\documentclass[12pt]{article}
\usepackage{graphicx}

\def\Kenensaw{Department of Physics\\
Kennesaw State University, Kennesaw, GA 30144, USA }
\def\support{\footnote{This material is based upon work supported by the National Science Foundation under Grant No. PHY 1519606.}}

\def\Title#1{\begin{center} {\Large #1 } \end{center}}
\def\Author#1{\begin{center}{ \sc #1} \end{center}}
\def\Address#1{\begin{center}{ \it #1} \end{center}}

\newenvironment{Abstract}{\begin{quotation}  }{\end{quotation}}
\newenvironment{Presented}{\begin{quotation} \begin{center} 
             PRESENTED AT\end{center}\bigskip 
      \begin{center}\begin{large}}{\end{large}\end{center} \end{quotation}}

\def\beq{\begin{equation}}
\def\eeq#1{\label{#1}\end{equation}}
\def\eeqn{\end{equation}}

\def\beqa{\begin{eqnarray}}
\def\eeqa#1{\label{#1}\end{eqnarray}}
\def\eeqan{\end{eqnarray}}

\let\bar=\overbar

\def\Dslash{\not{\hbox{\kern-4pt $D$}}}
\def\dslash{\not{\hbox{\kern-2pt $\del$}}}

\def\msb{{\bar{\ssstyle M \kern -1pt S}}}

\def\beq{\begin{equation}}
\def\eeq{\end{equation}}
\def\beqa{\begin{eqnarray}}
\def\eeqa{\end{eqnarray}}

\begin{document}
\begin{titlepage}

\vfill
\Title{Single-top production in the Standard Model and beyond}
\vfill
\Author{Nikolaos Kidonakis\support}
\Address{\Kenensaw}
\vfill
\begin{Abstract}
I present high-order calculations for single-top production in the Standard Model and in models with anomalous top-quark couplings. Theoretical results are presented for total cross sections and top-quark transverse momentum and rapidity distributions for the $t$ and $s$ single-top channels as well as for the associated production of a top quark with a $W$ boson in the Standard Model. Corrections from soft-gluon emission though NNNLO are included. I also show results for the associated production of a top quark with a $Z$ boson in processes involving anomalous top-quark couplings.
\end{Abstract}
\vfill
\begin{Presented}
CIPANP2018\\
Palm Springs, California, May 29--June 3, 2018
\end{Presented}
\vfill
\end{titlepage}
\def\thefootnote{\fnsymbol{footnote}}
\setcounter{footnote}{0}

\section{Higher-order soft-gluon corrections}

It has been well known for some time that higher-order QCD corrections 
are significant in top-quark production processes. 
Fixed-order results for single-top production are known at NLO \cite{BWHL,Zhu} 
and some at NNLO \cite{NNLOtch,BGYZ,BGZ,NNLOsch}. 

For single-top \cite{NKsingletop,NKtW16,NKtop} (as well as top-antitop \cite{NKtop,N3LOtt}) processes at Tevatron and LHC energies, it is also well-known that soft-gluon corrections are important, and that they approximate exact results very well to the order in which the latter are known. Further higher-order soft-gluon corrections provide an additional theoretical improvement. Here I provide results with soft-gluon corrections for $t$-channel and $s$-channel single-top production, and  
$tW$ production \cite{NKsingletop,NKtW16,NKtop}. I also present results for $tZ$ production via anomalous top-quark couplings \cite{NKtZ}.

For partonic processes of the form 
\beq
f_{1}(p_1)\, + \, f_{2}\, (p_2) \rightarrow t(p_t)\, + \, X 
\eeq
we define
$s=(p_1+p_2)^2$, $t=(p_1-p_t)^2$, $u=(p_2-p_t)^2$
and $s_4=s+t+u-\sum m^2$.
At partonic threshold $s_4 \rightarrow 0$.
Soft-gluon corrections appear in the perturbative corrections 
as plus distributions involving logarithms of $s_4$, i.e. 
$[\ln^k(s_4/m_t^2)/s_4]_+$.

\begin{figure}[htb]
\centering
\includegraphics[height=0.88in]{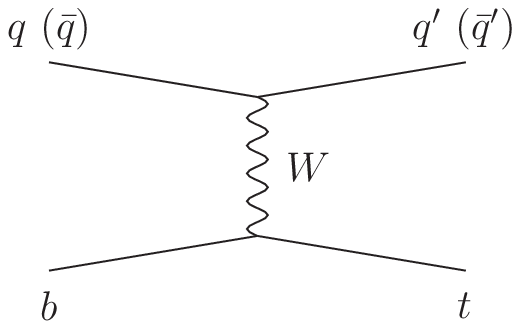}
\hspace{2mm}
\includegraphics[height=0.88in]{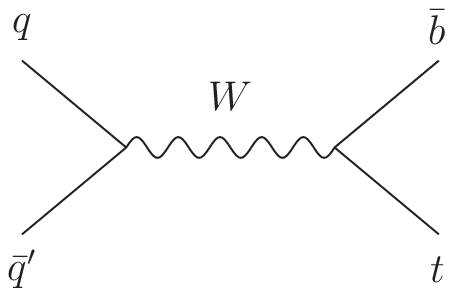}
\hspace{2mm}
\includegraphics[height=0.88in]{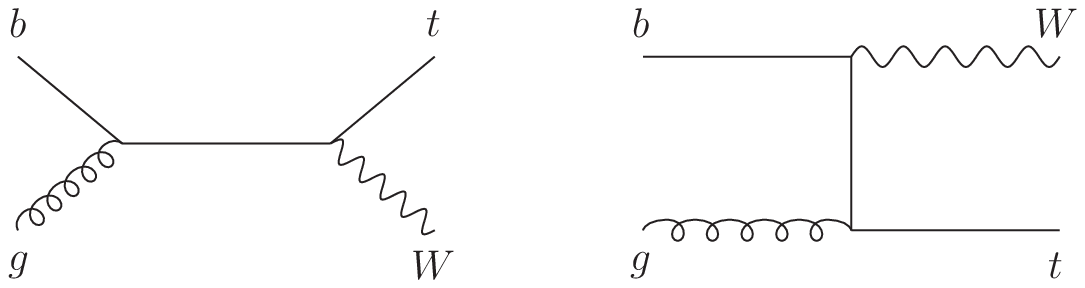}
\caption{Lowest-order diagrams for single-top $t$-channel (left), $s$-channel (second from left), and $tW$ (right two 
diagrams) production.}
\label{lodiag}
\end{figure}

We resum these soft-gluon corrections for the double-differential cross section at NNLL accuracy, which requires the calculation of two-loop soft anomalous dimensions \cite{NKsingletop,NK2loop,NK3loop}.
Taking moments of the partonic cross section with moment variable $N$,
${\hat \sigma}(N)=\int (ds_4/s) \; e^{-N s_4/s} {\hat \sigma}(s_4)$, 
we write a factorized expression for the cross section
\beqa
\sigma^{f_1 f_2\rightarrow tX}(N,\epsilon)
&=& H_{IL}^{f_1 f_2\rightarrow tX}\left(\alpha_s(\mu_R)\right) \, 
S_{LI}^{f_1 f_2 \rightarrow tX}\left(\frac{m_t}{N \mu_F},\alpha_s(\mu_R) \right)
\nonumber \\ && 
\times \prod  J_{\rm in} \left(N,\mu_F,\epsilon \right)
\prod J_{\rm out} \left(N,\mu_F,\epsilon \right)
\eeqa
where
$H_{IL}^{f_1 f_2\rightarrow tX}$ is a hard-scattering function  
and $S_{LI}^{f_1 f_2\rightarrow tX}$ is a soft-gluon function.

The evolution of the soft function,  
which gives the exponentiation of logarithms of $N$, 
follows from the renormalization group equation
\beq
\left(\mu \frac{\partial}{\partial \mu}
+\beta(g_s)\frac{\partial}{\partial g_s}\right)\,S_{LI}
=-(\Gamma^\dagger_S)_{LK}S_{KI}-S_{LK}(\Gamma_S)_{KI}
\eeq
where $\Gamma_S$ is the (process-dependent) soft anomalous dimension.
The resummed cross section may be written as a product of exponentials, 
involving universal terms describing collinear and soft-gluon emission from 
incoming and outgoing partons, as well as process-specific terms that describe 
noncollinear soft-gluon emission and that depend explicitly on the process and 
its color structure.

We use the resummed cross section as a generator of finite-order expansions, 
and we provide approximate NNLO (aNNLO) and approximate N$^3$LO (aN$^3$LO) 
predictions for cross sections and differential distributions.

\section{$t$-channel production}

\begin{figure}[htb]
\centering
\includegraphics[height=3.5in]{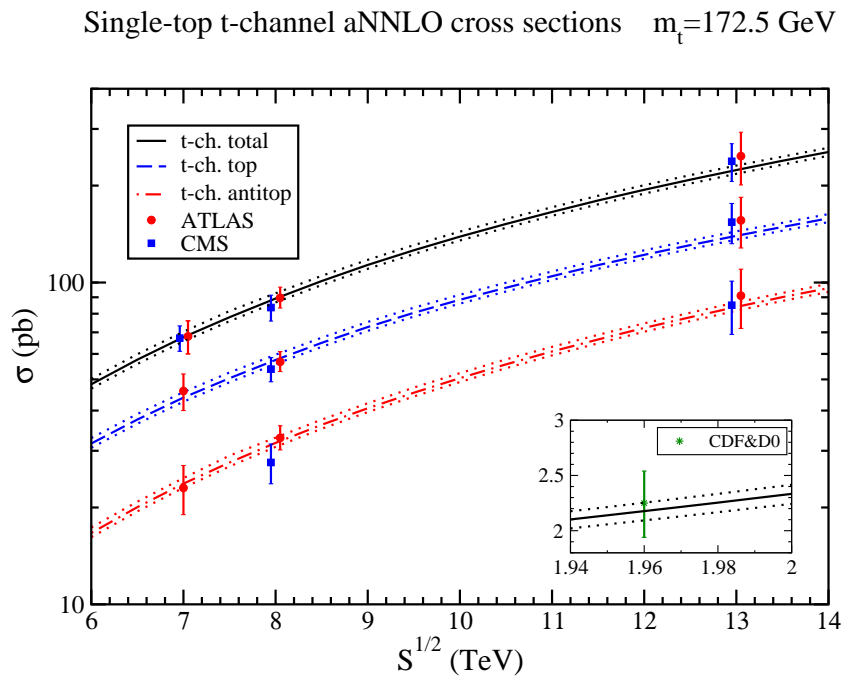}
\caption{Single-top aNNLO $t$-channel cross sections compared with CMS and ATLAS data at 7 TeV \cite{CMStch7,ATLAStch7}, 8 TeV \cite{CMStch8,ATLAStch8}, and 13 TeV \cite{ATLAStch13,CMStch13}, and (inset) with CDF and D0 combination data at 1.96 TeV \cite{CDFD0tch}.}
\label{tchplot}
\end{figure}

We begin with single-top production in the $t$-channel. The leading-order diagram is shown in Fig. \ref{lodiag}.

The soft anomalous dimension for this process is a $2\times 2$ matrix whose elements in a $t$-channel singlet-octet color basis at one loop in Feynman gauge are 
\beqa
{\Gamma}_{S\, 11}^{t \, (1)}&=&
C_F \left[\ln\left(\frac{t(t-m_t^2)}{m_t s^{3/2}}\right)-\frac{1}{2}\right] \, ,
\nonumber \\
{\Gamma}_{S\, 12}^{t \, (1)}&=&\frac{C_F}{2N} \ln\left(\frac{u(u-m_t^2)}{s(s-m_t^2)}\right) \, ,
\quad \quad 
{\Gamma}_{S\, 21}^{t \, (1)}= \ln\left(\frac{u(u-m_t^2)}{s(s-m_t^2)}\right) \, ,
\nonumber \\ 
{\Gamma}_{S\, 22}^{t \, (1)}&=& C_F \ln\left(\frac{s-m_t^2}{m_t \sqrt{s}}\right)
-\frac{1}{2N}\ln\left(\frac{t(t-m_t^2)}{s(s-m_t^2)}\right) 
+\frac{(N^2-2)}{2N}\ln\left(\frac{u(u-m_t^2)}{s(s-m_t^2)}\right)-\frac{C_F}{2} \, .
\nonumber \\ 
\eeqa

At two loops, we only need the first matrix element:
\beq
\Gamma_{S\, 11}^{t \, (2)}=\left[C_A\left(\frac{67}{36}-\frac{\zeta_2}{2}\right)
-\frac{5}{18}n_f\right] \Gamma_{S\, 11}^{t \, (1)}
+C_F C_A \frac{(1-\zeta_3)}{4} \, .
\eeq

We next provide numerical results for $t$-channel production at aNNLO. We use MMHT 2014 pdf \cite{MMHT2014}. Figure \ref{tchplot} shows the theoretical aNNLO $t$-channel production cross sections for single top, single antitop, and their sum, compared with data from the LHC and the Tevatron. Excellent agreement is observed between theory and data in all cases.

\begin{figure}[htb]
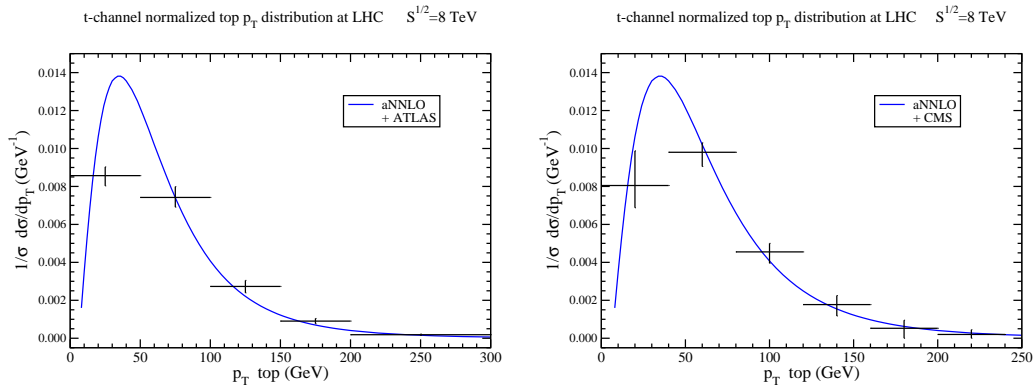

\centering
\includegraphics[height=2.0in]{ptnormtoptch8lhcATLAS1702.02859plot.eps}
\hspace{2mm}
\includegraphics[height=2.0in]{ptnormtoptch8lhcCMS-TOP-14-004plot.eps}
\caption{The aNNLO top-quark normalized $p_T$ distributions in $t$-channel production at 8 TeV compared to (left) ATLAS \cite{ATLAStch8} and (right) CMS \cite{CMStchpt8} data.}
\label{tchptnormplot}
\end{figure}

The aNNLO top-quark normalized $p_T$ distributions in $t$-channel production at 8 TeV energy are shown in Fig. \ref{tchptnormplot} and compared with LHC data; we again note the very good description of the data by the theory curves.

\section{$s$-channel production}

\begin{figure}[htb]
\centering
\includegraphics[height=3.5in]{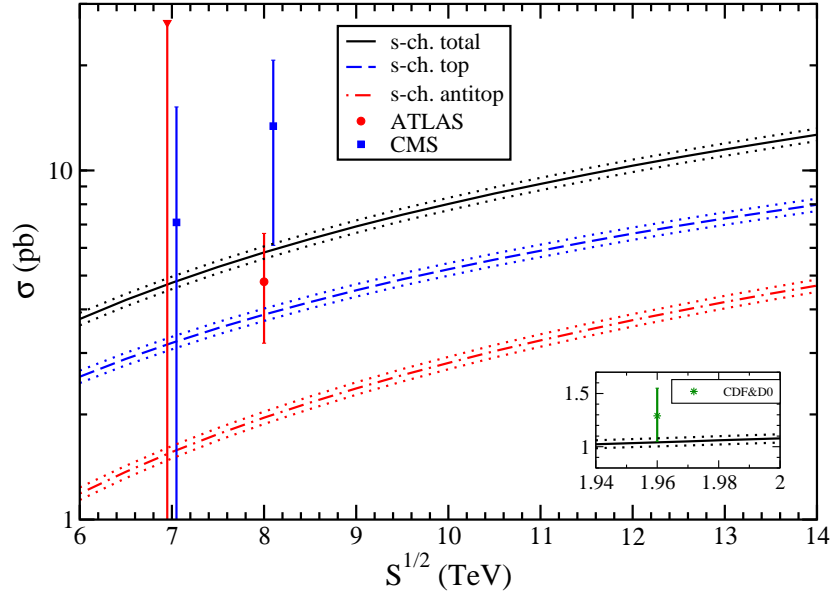}
\caption{Single-top aNNLO $s$-channel cross sections  
compared to ATLAS and CMS data at 7 TeV \cite{ATLASsch7,CMSsch7and8} and 
8 TeV \cite{CMSsch7and8,ATLASsch8}, and (inset) to CDF and D0 combination 
data \cite{CDFD0sch} at 1.96 TeV.}
\label{schplot}
\end{figure}

We continue with single-top production in the $s$-channel. The leading-order diagram is shown in Fig. \ref{lodiag}.

The soft anomalous dimension for this process is a $2\times 2$ matrix whose elements in an $s$-channel singlet-octet color basis at one loop in Feynman gauge are 
\beqa
\Gamma_{S\, 11}^{s \, (1)}&=&C_F \left[\ln\left(\frac{s-m_t^2}{m_t\sqrt{s}}\right)
-\frac{1}{2}\right] \, ,
\nonumber \\ 
\Gamma_{S\, 12}^{s \, (1)}&=&\frac{C_F}{2N} \ln\left(\frac{u(u-m_t^2)}{t(t-m_t^2)}\right) \, ,
\quad \quad
\Gamma_{S\, 21}^{s \, (1)}= \ln\left(\frac{u(u-m_t^2)}{t(t-m_t^2)}\right) \, , 
\nonumber \\ 
\Gamma_{S\, 22}^{s \, (1)}&=&C_F \ln\left(\frac{s-m_t^2}{m_t \sqrt{s}}\right)
-\frac{1}{N}\ln\left(\frac{u(u-m_t^2)}{t(t-m_t^2)}\right)
+\frac{N}{2} \ln\left(\frac{u(u-m_t^2)}{s(s-m_t^2)}\right)-\frac{C_F}{2} \, .
\eeqa

At two loops, we only need the first matrix element:
\beq
\Gamma_{S\, 11}^{s\, (2)}=\left[C_A\left(\frac{67}{36}-\frac{\zeta_2}{2}\right)
-\frac{5}{18}n_f\right] \Gamma_{S\, 11}^{s\, (1)}
+C_F C_A \frac{(1-\zeta_3)}{4} \, .
\eeq

Figure \ref{schplot} shows the theoretical aNNLO $s$-channel production cross sections, using MMHT 2014 pdf \cite{MMHT2014}, for single top, single antitop, and their sum, compared with available data from the LHC and the Tevatron.

\section{Associated $tW$ production}

\begin{figure}[htb]
\centering
\includegraphics[height=3.5in]{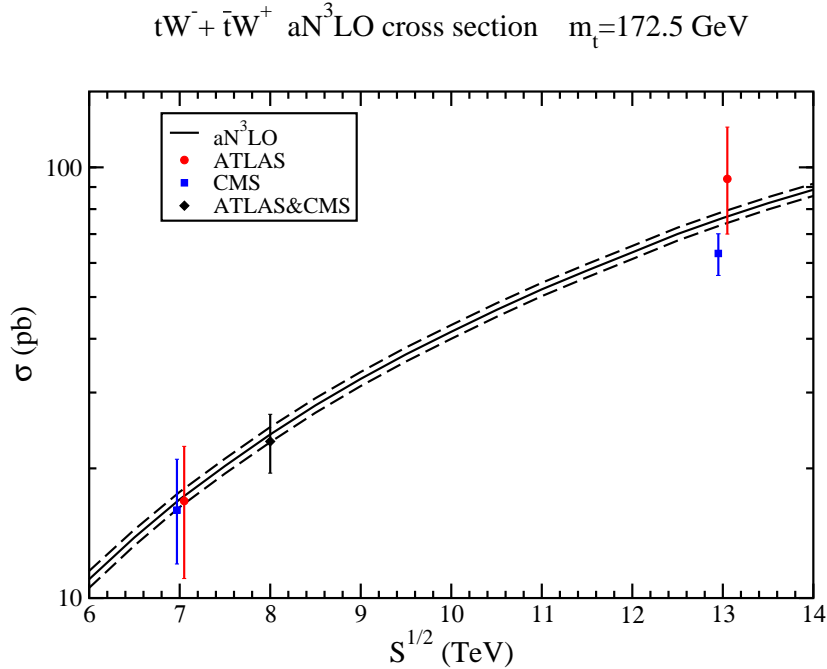}
\caption{Single-top aN$^3$LO cross sections for $tW$ production 
compared to ATLAS and CMS data at 7 TeV \cite{ATLAStW7,CMStW7}, 8 TeV \cite{ATLASCMStW8}, and 13 TeV \cite{ATLAStW13,CMStW13}.}
\label{tWplot}
\end{figure}

\begin{figure}[htb]
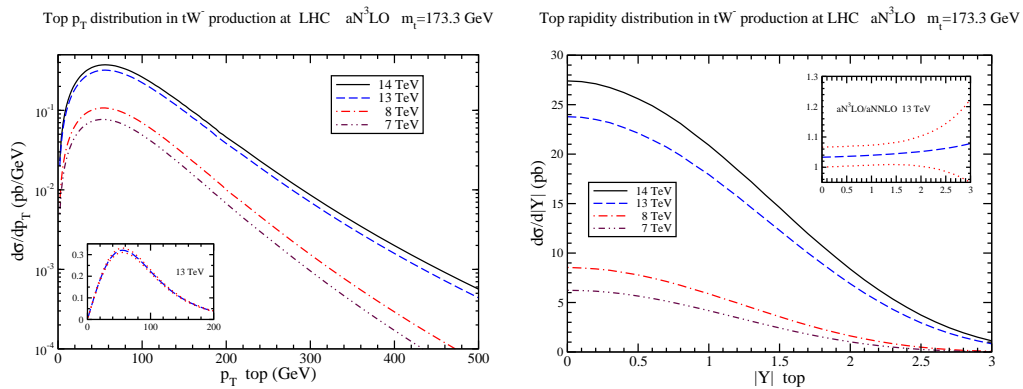

\centering
\includegraphics[height=2.0in]{pttoptWplot.eps}
\hspace{2mm}
\includegraphics[height=2.0in]{yabstoptWplot.eps}
\caption{Top-quark aN$^3$LO differential distributions in (left) $p_T$ and (right) rapidity in $tW^-$ production.}
\label{tWptyplot}
\end{figure}

We continue with single-top production in association with a $W$ boson. The leading-order diagrams for $tW$ production 
are shown in Fig. \ref{lodiag}.

The soft anomalous dimension for this process at one loop in Feynman gauge is
\beq
\Gamma_S^{tW^- \, (1)}=C_F \left[\ln\left(\frac{m_t^2-t}{m_t\sqrt{s}}\right)
-\frac{1}{2}\right] +\frac{C_A}{2} \ln\left(\frac{m_t^2-u}{m_t^2-t}\right) \, .
\eeq

At two loops, we have
\beq
\Gamma_S^{tW^- \, (2)}=\left[C_A\left(\frac{67}{36}-\frac{\zeta_2}{2}\right)
-\frac{5}{18}n_f\right] \Gamma_S^{tW^- \, (1)}
+C_F C_A \frac{(1-\zeta_3)}{4} \, .
\eeq

Figure \ref{tWplot} shows the theoretical aN$^3$LO results for the sum of the $tW^-$ and ${\bar t}W^+$ cross sections, using MMHT 2014 pdf \cite{MMHT2014}, compared with data from the LHC.
Excellent agreement is found between theory and data for all LHC energies.
The soft-gluon corrections are large; we note that large corrections are also found in related $W$ and $Z$ production processes at high $p_T$ \cite{WZ}.

Figure \ref{tWptyplot} shows the aN$^3$LO top-quark $p_T$ and rapidity distributions in $tW$ production. As the inset plot on the right shows, the aN$^3$LO corrections are significant and increase at higher rapidities. 

\section{$tZ$ production via anomalous couplings}

\begin{figure}[htb]
\centering
\includegraphics[height=1.0in]{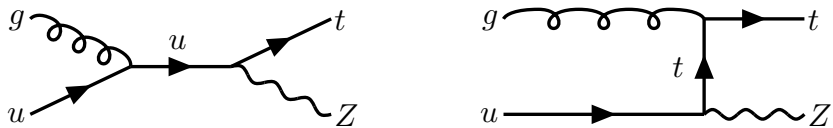}
\caption{Lowest-order diagrams fot $tZ$ production via anomalous couplings.}
\label{tZplot}
\end{figure}

In addition to Standard Model processes for top production, top quarks may also be produced via anomalous top-quark couplings \cite{NKtZ,fcnc}. Here we study $tZ$ production via such couplings \cite{NKtZ}. A Lagrangian for such processes is given by
\beq
\Delta {\cal L}^{eff} =    \frac{1}{ \Lambda } \,
\kappa_{tqZ} \, e \, \bar t \, (i/2) (\gamma_{\mu}\gamma_{\nu}
-\gamma_{\nu}\gamma_{\mu}) \, q \, F^{\mu\nu}_Z + h.c.
\eeq
Figure \ref{tZplot} shows the leading-order diagrams for the process $gu \rightarrow tZ$; the similar process $g c \rightarrow tZ$ also contributes.

The NLO corrections to these processes were calculated in Ref. \cite{NLOtqZ}, and we find that they are dominated by soft gluons. In our numerical results below we use CT14 \cite{CT14} pdf.

\begin{figure}[htb]
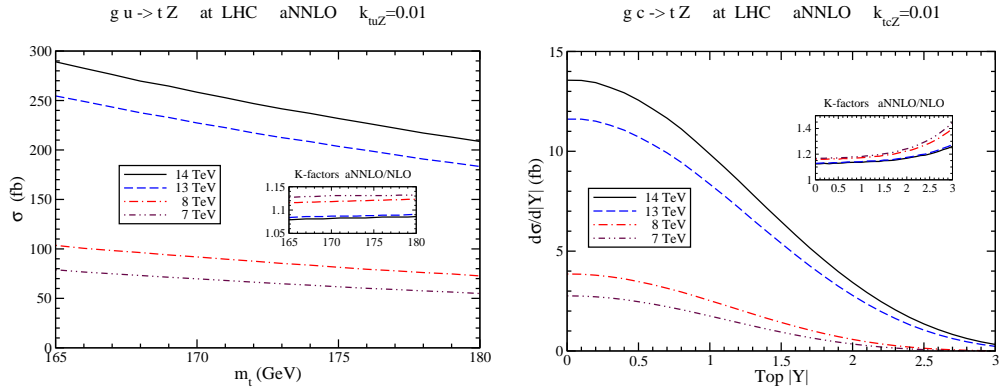

\centering
\includegraphics[height=2.0in]{gutZlhcv2plot.eps}
\hspace{2mm}
\includegraphics[height=2.0in]{yabstopgctZlhcv2plot.eps}
\caption{(Left) Total cross sections at aNNLO \cite{NKtZ} for $tZ$ production via $gu\rightarrow tZ$; 
(Right) Top-quark rapidity distributions at aNNLO for $tZ$ production via $gc\rightarrow tZ$.}
\label{gutZgctZ}
\end{figure}

In Fig. \ref{gutZgctZ} we show aNNLO results for $tZ$ production. The left plot shows total cross sections at LHC energies for $gu \rightarrow tZ$ while the right plot shows the top-quark rapidity distributions for $gc \rightarrow tZ$.
The insets in both plots show the aNNLO/NLO ratios, which demonstrate that the 
higher-order soft-gluon corrections are large, especially at large rapidities.  

These theoretical results may be useful in setting limits on the couplings, which is an area of active study by the LHC experiments \cite{CMStqz,ATLAStqz}.

\section{Summary}

I have presented cross sections and differential distributions for single-top production. Results with soft-gluon corrections at aNNLO for $t$-channel and $s$-channel production, and at aN$^3$LO for $tW$ production, have been presented for total cross sections and top-quark $p_T$ and rapidity distributions. The soft-gluon corrections are significant and we find excellent agreement of the theory with LHC and Tevatron data.

I have also presented aNNLO results for $tZ$ production via anomalous top-quark couplings. Soft-gluon corrections are also very significant for this process.

\end{document}